          [2001/04/25 1.1 (PWD)]
\documentclass[a4paper,twocolumn]{esapub} 
\usepackage{natbib,graphicx}

\title{Kink waves in solar spicules: observation and modelling}
\author{V. Kukhianidze$^1$}
\author{T.V. Zaqarashvili$^{1,2}$}
\author{E. Khutsishvili$^1$}
\affil{\it 1. Georgian National Astrophysical Observatory (Abastumani
Astrophysical Observatory), Al. Kazbegi ave. 2a, 0160 Tbilisi,
Georgia, E-mail: vaso@genao.org, temury@genao.org, eldar@genao.org\\
\it 2. Departament de F\'{\i}sica,
Universitat de les Illes Balears, E-07122 Palma de Mallorca, Spain, E-mail: temury.zaqarashvili@uib.es}

\begin{document}

\keywords{solar atmosphere; kink waves}

\maketitle

\begin{abstract}
Height series of Doppler observation at the solar limb (covering 3800 - 8700 km distance above the photosphere)
in $H_{\alpha}$ spectral line obtained by big coronagraph of Abastumani Astrophysical Observatory \citep{khu} show 
the periodic spatial distribution of Doppler velocities in spicules. 
We suggest that the periodic spatial distribution is caused by propagating kink waves in spicule. The wave length is found to be $\sim$ 3500 km. Numerical modelling of kink wave propagation from the photosphere to observed heights gives the wave length of kink waves at the photosphere to be $\sim$ 1000 km,  which indicates to the granular origin of the waves. The period of waves is estimated to be in the range of 35-70 s.

\end{abstract}

\section{Introduction}

Most of chromospheric radiation in quiet Sun regions comes from
spicules, which are jet-like chromospheric structures observed at
the solar limb mainly in $H_{\alpha}$ line. Spicules usually show
the group behaviour and probably are concentrated between
supergranule cells (see e.g. reviews of \citet{bec} and
\citet{ste}), thus their formation is connected to 
intense magnetic fields. Typical life time of spicules is 5-15 minutes. 
Therefore the propagation of MHD waves through the chromosphere with shorter period than the 
spicule lifetime can be traced through spicule observations. 

More then 30 years ago \citet{nik} reported that spicules oscillate
along the limb with characteristic period of about 1 minute and
velocities about 10-15 km/s. The oscillation, which they reported, 
was just periodic transversal displacement of spicule axis at one 
particular height from the photosphere. 

If spicules are formed in thin magnetic flux tubes, then the periodic transverse displacement 
of the axis observed by \citet{nik} probably is due to
the propagation of kink waves. The kink waves can be generated in photospheric tubes
by buffeting of granular motions \citep{rob0,hol,spru,has}. As kink waves cause the displacement of tube axis, then their propagation can be revealed either by direct observation of the periodic displacement of 
spicule axis along the limb as in \citet{nik} or by the Doppler
shift of spectral line. The later possibility arises when the kink oscillation occurs in plane of observations.

The periodic spatial distribution of Doppler velocities in spicules has been 
observed almost 20 years ago by \citet{khu} (hereinafter paper I). Unfortunately, neither 
observers nor theorists paid attention to the phenomenon which can be simply explained
in terms of kink waves. In this paper, we model the observations as propagating kink 
waves excited due to the photospheric granulation.

\section[]{Observation}

The big 53 cm coronagraph and universal spectrograph of Abastumani Astrophysical Observatory have been
used to obtain chromospheric $H_{\alpha}$ and $D_3$ line spectra
at different heights (8 heights) from the photosphere (see Paper I). 
Instrumental spectral resolution is 0.04 {\AA} and dispersion is 1 {\AA}/mm in $H_{\alpha}$. 
Observations were carried out at the solar limb as height series beginning at 3800 km height from the photosphere and
upwards in coronal days when spatial resolution was 1 arc sec (see paper I). 
The distance between neighboring heights was 1 arc sec, thus the spatial distance
3800-8700 km above the photosphere was covered. Here we only consider the $H_{\alpha}$ observations, however $D_3$ line spectra are also 
interesting and can be studied in future.
The exposure time was 0.4 s at lower heights and 0.8 at higher ones for
$H_{\alpha}$. The total time duration of one
height series was 7 s. The consequent height
series begins immediately. The total duration of the observation
was 44 min. Therefore at one height we have a continuous
$H_{\alpha}$ spectra with time intervals of ~7 s. Approximately
300 height series was taken. More details about the observations can 
be found in the paper I. 

\begin{figure}
\centering
\includegraphics[width=0.9\linewidth]{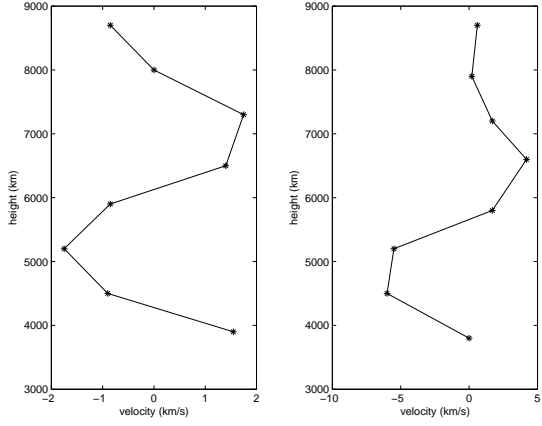}
\caption{Two height series of Doppler velocity in two different
spicules in $H_{\alpha}$ spectral line are shown. Marked dots indicate to the observed heights 
(note that time difference between consecutive height observation is 1 s).
The velocity has clear periodic spatial distribution, which can be caused by the propagation of kink waves. The wave lengths are around $3500$ km.\label{fig:single}}
\end{figure}

We analayzed measured spatial distributions of Doppler velocities in $H_{\alpha}$ 
presented in the Paper I. Nearly 20$\%$ of spicules show  periodic spatial distributions of Doppler
velocities in height series \citep{khu1}. The Doppler velocity spatial
distributions in two height series of two different spicules in
$H_{\alpha}$ spectral line are shown on Fig.1. The periodic
spatial distributions are clearly seen. So the transversal
velocity in that particular spicules is periodically redistributed
with height, which indicates to propagating or standing waves. The wave length of propagating waves is 

\begin{equation}
{\sim} 3500 \,\, km.
\end{equation}

The maximal time difference between lowest (3800 km above the photosphere) and highest (8700 km above the photosphere) 
heights is 7 s, which is the duration of each height series. Estimation of the distance at which the wave can travel during this time gives  $\sim$ 350 km (for the Alfv{\'e}n speed $\sim$ 50 km/s in the chromosphere), which is less than spatial resolution 
of the coronagraph ($\sim$ arc sec), thus less than distance between consequent observed heights. Therefore we may consider the velocity distribution 
during one height series as pictured at the same time. 

Thus observations show the evidence of transversal wave propagation in spicules. The waves can be either Alfv{\'e}n or kink waves.
But as spicules are highly structured phenomena (their width are $\sim$ 1 arc sec), the waves probably are kink waves. In next section we model the observations as propagating kink waves in spicules.

\section{Theoretical modelling}

The mechanism of spicule formation is not well understood, therefore we do not concern here how spicules have been formed (for excitation mechanisms see review of \citet{ste} and references therein). As expected kink wave period (with phase speed of 50 km/s and wave length of 3500 km the period is 70 s) 
is much shorter than the spicule life time, then during this short time interval we may consider the spicules as just existing stable structure. So we may study the wave propagation into this stable structure. Hence despite of very dynamic behaviour of spicules, the approximation allows to gain main properties of wave propagation.  

Spicules can be modelled as thin magnetic flux tubes embedded in field free environment, anchored in the photosphere and persisted
towards the corona. The equilibrium state of stratified atmosphere is defined by hydrodynamic vertical pressure balance  
\begin{equation}
{\partial p_e}/{\partial z} = - \rho_e(z)g,
\end{equation}
where $p_e(z), \rho_e(z)$ are the hydrodynamic pressure and the plasma density through the atmosphere and $g$ is the gravitational acceleration.
The equilibrium state of vertical thin magnetic flux tube also requires transversal pressure balance at tube boundaries 
 \begin{equation}
 p_0(z) + {B^2_0(z)}/{8\pi}=p_e(z)
 \end{equation} 
where $p_0(z)$ is the hydrodynamic pressure inside the tube and $B_0(z)$ is tube magnetic field.
 
Propagation of kink waves in vertical magnetic tube embedded in field free environment is governed by the equation \citep{rob0,spru,rob1}
\begin{equation}
{{\partial^2 {\xi}}\over {\partial t^2}}= c^2_k{{\partial^2
{\xi}}\over {\partial z^2}} + g{{\rho_0 - \rho_e}\over {\rho_0 +
\rho_e}}{{\partial {\xi}}\over {\partial z}},
\end{equation}
where $\xi$ is the transverse displacement, $\rho_0(z)$ is the plasma density inside the tube, 
$$c_k=c_A\sqrt{{{\rho_0}\over {\rho_0 +
\rho_e}}}
$$ is the kink speed and $$c_A={B_0\over {\sqrt {4\pi\rho_0}}}$$ is the Alfv{\'e}n speed.

Kink waves propagating along the magnetic tube lead to the oscillation of the tube axis (see Fig.2). This results in the periodic Doppler shift of observed spectral line (in our case $H_{\alpha}$) when the kink waves oscillate in the plane of observations. At 
the same time the Doppler shift will have periodic behaviour in height: it will be maximal at velocity antinodes (blue and red shifted at corresponding antinodes) and tends to zero at velocity nodes (for torsional Alfv{\'e}n waves see \citet{zaq}). So kink waves give the similar behaviour of Doppler shift as the observations (Fig.1). Therefore we argue that the propagating kink waves are the most plausible reason of observed spatial periodicity in $H_{\alpha}$ Doppler velocity.

\begin{figure}
\centering
\includegraphics[width=0.6\linewidth]{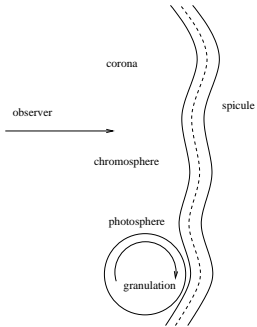}
\caption{Schematic picture of propagating kink waves in spicules. Due to the propagation of kink waves the spectral line will be Doppler shifted periodically in space if the wave velocity is polarised in the plane of observation. The Doppler shift will be maximal at velocity antinodes (blue and red shifted at corresponding antinodes) and zero at velocity nodes.
\label{fig:single}}
\end{figure}

Spicule density and magnetic field show almost no spatial variation at observed heights (3800-8700 km above the photosphere) and the stratification also can be neglected, then the equation (4) gives the approximate dispersion relation:
\begin{equation}
c^2_k k^2_z = \omega^2,
\end{equation}   
where $k_z$ is the vertical wave number and $\omega$ is the frequency of kink waves. From the
dispersion relation we may estimate the period of kink waves using the observed wave length and particular kink speed. The typical observed wave lengths are of order $\sim 3500$ km (see Fig.1). Then for the kink speed being as $\sim 50-100$ km/s the expected period of kink waves is in the range of 
\begin{equation}
{\sim} 35-70 \,\, s.
\end{equation}

The cut-off frequency of kink waves due to stratification at the photosphere is 
\begin{equation}
\omega^2_c = {{g}\over {2h(2\beta +1)}}\left [{1\over 4} +
{{\partial h}\over {\partial z}} \right ],
\end{equation}
where $h$ is pressure scale height. For an isothermal medium the pressure scale height at the photosphere is $\sim$ 125 km, which gives (with $\beta=1$) the cut-off period as $\sim$ 660 s. So the expected period of kink waves is well below the cut-off value. Thus the kink waves with the period of $\sim$ 35-70 s may easily propagate upwards.

Thus estimations show that observed kink waves have higher frequencies than the 5-minute oscillation in the photosphere. Therefore p-modes can be ruled out to be the source of the waves. This immediately turns to the granulation. The photospheric granulation has been suggested as the source of kink waves in thin tubes \citep{spru,hol,has}. Mean granular diameter is 800-1000 km, therefore the dynamic granular cell probably generates the kink waves with wave length similar to the diameter (see Fig.2). But the observed wave length of kink waves (3500 km) in higher atmosphere is few times longer than mean granular diameter. The discrepancy probably arises due to the increase of kink speed 
with height: granular cells generate the kink waves with the wave length comparable to their diameter, but when the wave propagate upwards the wave length increases due to the increasing phase speed. 

\begin{figure}
\centering
\includegraphics[width=0.9\linewidth]{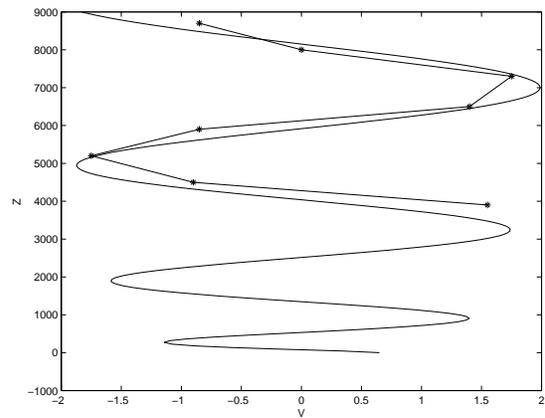}
\caption{The comparison of observations (first plot on Fig.1) and the numerical modelling. Vertical axis shows the height in km and the horizontal axis shows the kink wave velocity in km/s. It is seen that the computed kink wave is best fitted to the observation if its wave length at the photosphere is $\sim$ 1000 km. In numerical simulation the Alfv{\'e}n speed varies from 10 km/s at the photosphere up to 100 km/s at the height 6000 km. The period of waves is 40 s.
\label{fig:single}}
\end{figure}   

In order to test this assumption we should solve the equation (4) along whole solar atmosphere from the photosphere to observed heights ($\sim$ 4000-9000 km). However the height distribution of physical quantities are not well known. There are only suggested values of density and magnetic field in spicules and they are also controversial. Thus the difference between plasma densities inside and outside the spicule is not well determined. At the photospheric level the thermal equilibrium inside and outside the tube i.e. $T_0(z)=T_e(z)$ is a good approximation, which gives higher density outside the tube than inside (in order to held the equilibrium (2)). But in higher atmosphere, say at heights $>$ 2000 km, the plasma temperature is definitely lower in spicules than in environment. This is because the spicules are seen in chromospheric spectral lines formed at much lower temperature than the surrounding coronal plasma. Also the plasma density seems to be much higher in spicules than in surrounding coronal plasma. Due to this uncertainty of medium parameters, we take the height distribution of Alfv{\'e}n speed and the density ratio inside and outside spicule $\rho_0/\rho_e$ instead of separate height dependence of density and magnetic field. We take the value of $10$ km/s for the Alfv{\'e}n speed at the photosphere, which increases up to  $\sim 100$ km/s at the height 6000 km. Also we take 1/3 for the density ratio at the photosphere (being higher outside the tube), which increases up to 10$^3$ at the height 6000 km. Thus both, Alfv{\'e}n speed and density ratio are linear functions of height
\begin{equation}
c_A = 10 + 1.875{z\over h} \,\,\, km/s,
\end{equation}
\begin{equation}
{{\rho_0}\over {\rho_e}}= {1\over 3} + {62.5\over 3}{z\over h},
\end{equation}
where $h$=125 km is pressure scale height at the photosphere and $z=0$ corresponds to the solar surface.
             
Taking these values of Alfv{\'e}n speed and density ratio we solve numerically the kink wave equation for different wave frequency. And then we fitted the numerical solution to the observed curves of Doppler velocity. Result of numerical solution fitted to observed height series is plotted on the Fig.3. It is seen that the numerical solution is well fitted to the observation. In this case the wave period is taken as 
\begin{equation}
\sim 40 \,\,\, s.
\end{equation}

From Figure 3 it is seen that the wave length of kink waves at the photospheric level is 
\begin{equation}\sim 1000 \,\,\, km,
\end{equation}
which is comparable to the granular diameter. It clearly indicates to the granular origin of the waves. 

\section{Conclusions}

Height series of Doppler observation at the solar limb (3800 - 8700 km distance above the photosphere)
in $H_{\alpha}$ spectral line show the periodic spatial distribution of Doppler velocities in spicules. 
We suggest that the periodic spatial distributions are caused by propagating kink waves in spicules. The wave length at observed heights is $\sim$ 3500 km. Numerical modelling of kink wave propagation from the photosphere to the observed heights is best fitted to the observations when the wave period is $\sim$ 40 s. In this case the wave length of kink waves at the photosphere is $\sim$ 1000 km, which probably indicates to the granular origin of the waves. 

Our suggested scenario is the following: due to the nonstationary character of granular motion the kink waves with the wave length comparable to the granular diameter are generated in thin magnetic flux tubes at the granular boundaries. Then the waves propagate upwards along the magnetic flux tube towards the corona (see Fig.2) and cause the observed height variation of Doppler shift in spicules.

\section{Acknowledgements}

T. Zaqarashvili acknowledges the financial support from conference organisers.
Authors thank to Prof. B. Roberts for useful suggestions. 
The work is supported by the grant of Georgian Academy of Sciences. 
T.Z. was partially supported by the NATO Reintegration Grant FEL.RIG 980755.

\end{document}